\documentclass[conference]{IEEEtran}
\IEEEoverridecommandlockouts
\usepackage{cite}
\usepackage{amsmath,amssymb,amsfonts}
\usepackage{algorithmic}
\usepackage{graphicx}
\usepackage{textcomp}
\usepackage{xcolor}
\definecolor{lightgray204}{RGB}{204,204,204}
\usepackage{comment}
\usepackage{soul}
\sethlcolor{red}
\usepackage{url}
\usepackage[acronym,shortcuts,nonumberlist,nopostdot,nomain,nogroupskip]{glossaries}
\newacronym{CSI-RS}{CSI-RS}{channel state information-reference signal}
\newacronym{SRS}{SRS}{sounding reference signal}
\newacronym{P2P}{P2P}{point to point}
\newacronym{UPA}{UPA}{uniform planar array}
\newacronym{ULA}{ULA}{uniform linear array}
\newacronym{PMI}{PMI}{precoding matrix indicator}
\newacronym{RI}{RI}{rank indicator}
\newacronym{CQI}{CQI}{channel quality indicator}
\newacronym{DL}{DL}{downlink}
\newacronym{CDF}{CDF}{cumulative distribution function}
\newacronym{DFT}{DFT}{discrete Fourier transform}
\newacronym{2D-DFT}{2D-DFT}{two-dimensional discrete Fourier transform}
\newacronym{NR}{NR}{new radio}
\newacronym{LTE}{LTE}{long term evolution}
\newacronym{PL}{PL}{path loss}
\newacronym{AF}{AF}{array factor}
\newacronym{AH}{AH}{aerial highway}
\newacronym{PSO}{PSO}{particle swarm optimization}
\newacronym{eGA}{eGA}{elite genetic algorithm}
\newacronym{GA}{GA}{genetic algorithm}
\newacronym{ICC}{ICC}{international conference on communications}
\newacronym{SSB}{SSB}{synchronization signal block}
\newacronym{ES}{ES}{Eigenscore}
\newacronym{AoA}{AoA}{angle of arrival}
\newacronym{AoD}{AoD}{angle of departure}
\newacronym{UAV}{UAV}{unmanned aerial vehicle}
\newacronym{CCUAV}{CCUAV}{cellular connected unmanned aerial vehicle}
\newacronym{D2D}{D2D}{device to device}
\newacronym{gUE}{gUE}{ground user equipment}
\newacronym[
    plural={UEs},
    longplural={user equipment}
]{UE}{UE}{user equipment}
\newacronym{MIMO}{MIMO}{multiple-input multiple-output}
\newacronym{mMIMO}{mMIMO}{massive multiple-input multiple-output}
\newacronym{MU-mMIMO}{MU-mMIMO}{multi-user massive multiple-input multiple-output}
\newacronym{MU-MIMO}{MU-MIMO}{multi-user multiple-input multiple-output}
\newacronym{SU-MIMO}{SU-MIMO}{single-user multiple-input multiple-output}
\newacronym{PRB}{PRB}{physical resource block}
\newacronym{RE}{RE}{resource element}
\newacronym{RSRP}{RSRP}{reference signal received power}
\newacronym{RSS}{RSS}{received signal strength}
\newacronym{mmWave}{mmWave}{millimetre wave}
\newacronym{eICIC}{eICIC}{enhanced inter-cell interference coordination}
\newacronym{SINR}{SINR}{signal-to-interference-plus-noise ratio}
\newacronym{UAM}{UAM}{urban air mobility}
\newacronym{QoS}{QoS}{quality of service}
\newacronym{LoS}{LoS}{line of sight}
\newacronym{NLoS}{NLoS}{not line of sight}
\newacronym{BVLoS}{BVLoS}{beyond visual line of sight}
\newacronym{DoF}{DoF}{degree of freedom}
\newacronym{ZF}{ZF}{zero forcing}
\newacronym{CSI}{CSI}{channel state information}
\newacronym{3GPP}{3GPP}{third generation partnership project}
\newacronym{SVD}{SVD}{single value decomposition}
\newacronym{PBCH}{PBCH}{physical broadcast channel}
\newacronym{PDSCH}{PDSCH}{physical data shared channel}
\newacronym{thp}{thp}{throughput}
\newacronym{UMi}{UMi}{urban micro}
\newacronym{UMa}{UMa}{urban macro}
\newacronym{CAGR}{CAGR}{compound annual growth rate}
\newacronym{HO}{HO}{handover}
\newacronym{MNO}{MNO}{mobile network operator}
\newacronym{NOMA}{NOMA}{non-orthogonal multiple access}
\newacronym{BO}{BO}{bayesan optimization}
\newacronym{ML}{ML}{machine learning}
\newacronym{FR1}{FR1}{frequency range 1}
\newacronym{SO}{SO}{southern}
\newacronym{E}{E}{eastern}
\newacronym{NE}{NE}{northeastern}
\newacronym{RAN}{RAN}{radio access network}
\newacronym{BS}{BS}{base station}
\newacronym{ISD}{ISD}{inter-site distance}
\newacronym{IGD}{IGD}{inter-grid distance}
\newacronym{IHD}{IHD}{inter-highway distance}
\newacronym{IUD}{IUD}{inter-UAV distance}
\newacronym{RRC}{RRC}{radio resource control}
\newacronym{PSS}{PSS}{primary synchronization signal}
\newacronym{SSS}{SSS}{secondary synchronization signal}
\newacronym{MINP}{MINP}{mixed-integer nonlinear problem}
\newacronym{PAHSS}{PAHSS}{Particle Aerial Highway Swarm Segmentation}
\newacronym{MAMA}{\texttt{MAMA}}{mMIMO-Aerial-Metric-Association}
\newacronym{URD}{URD}{urban random distributed}
\newacronym{UDN}{UDN}{ultra dense network}
\newacronym{ADAM}{ADAM}{ADAptive Moment estimation}
\newacronym{CRS}{CRS}{common reference signal}
\newacronym{MRT}{MRT}{maximum ratio transmission}
\newacronym{rhs}{rhs}{right hand side}
\newacronym{lhs}{lhs}{left hand side}
\newacronym{IoT}{IoT}{internet of thing}
\newacronym{KPI}{KPI}{key performance indicator}
\newacronym{NTN}{NTN}{non-terrestrial network}
\newacronym{TN}{TN}{terrestrial network}
\newacronym{HAPS}{HAPS}{high-altitude platform station}
\newacronym{CF-mMIMO}{CF-mMIMO}{cell free mMIMO}
\newacronym{EASA}{EASA}{European Union Aviation Safety Agency}
\newacronym{C2}{C2}{Command and Control}
\newacronym{UTM}{UTM}{unmanned aircraft system traffic management}

\newacronym{EE}{EE}{Energy Efficient}
\newacronym{SE}{SE}{Spectral Efficient}
\newacronym{UPV}{UPV}{Universitat Politecnica de Valencia}
\newacronym{RS}{RS}{reference signal}
\newacronym{PCI}{PCI}{physical cell identity}
\newacronym{RTT}{RTT}{round trip time}

\newacronym{SORA}{SORA}{Specific Operations Risk Assessment}

\newacronym{DAA}{DAA}{Detect and Avoid}
\newacronym{USSP}{USSP}{U-Space Service Provider}

\newacronym{SLA}{SLA}{service level agreement}

\newacronym{QoM}{QoM}{quality of mission}
\newacronym{QoE}{QoE}{quality of experience}
\newacronym{XR}{XR}{extended reality}

\newacronym{ISAC}{ISAC}{Integrated sensing and communication}

\newacronym{DT}{DT}{digital twin}
\newacronym{RRM}{RRM}{radio resource management}

\newacronym{ICT}{ICT}{information and communication technology}
\newacronym{RU}{RU}{radio unit}
\newacronym[plural=RATs,
            longplural=radio access technologies]
            {RAT}{RAT}{radio access technology}

\newacronym{BBU}{BBU}{baseband unit}
\newacronym{MCPA}{MCPA}{multi-carrier power amplifier}

\newacronym{CS}{CS}{carrier shutdown}

\newacronym{RL}{RL}{reinforcement learning}
\newacronym{LEO}{LEO}{low Earth orbit}

\newacronym{NGMN}{NGMN}{Next Generation Mobile Networks Alliance}

\setglossarystyle{list} 

\usepackage[normalem]{ulem}

\usepackage[caption=false,font=small,labelfont=sf,textfont=sf]{subfig} % for subfigure
\usepackage{float}
\usepackage{tikz}
\usepackage{pgfplots}
\usetikzlibrary{patterns}
\pgfplotsset{compat=newest}
\newlength\fheight
\newlength\fwidth

\def\BibTeX{{\rm B\kern-.05em{\sc i\kern-.025em b}\kern-.08em
    T\kern-.1667em\lower.7ex\hbox{E}\kern-.125emX}}

\DeclareMathOperator{\lognormal}{LogNormal}

\renewcommand{\figurename}{Figure}

\begin{document}

% \title{HAPS-Assisted Energy-Efficient Carrier Management for Dense Urban 6G Non-Terrestrial Networks}
\title{HAPS as a Hypercell: Enabling Coverage and Capacity Carrier  Shutdown in Cellular Networks}

\author{
% \vspace{-0.3cm}
\IEEEauthorblockN{
Matteo Bernabè\IEEEauthorrefmark{1}, 
David López-Pérez\IEEEauthorrefmark{1}
and 
Nicola Piovesan\IEEEauthorrefmark{7}
}
% \\ %\vspace{-0.1cm}
\normalsize\IEEEauthorblockA{\emph{
\IEEEauthorrefmark{1}Universitat Politècnica de València (UPV), Spain} 
% \emph{
% \IEEEauthorrefmark{2}Beihang Valencia Polytechnic Institute (BVPI), China} 
}
\normalsize\IEEEauthorblockA{\emph{
\IEEEauthorrefmark{7}Huawei Technologies, France} 
}

\thanks{
This research is supported by the Generalitat Valenciana, Spain, through the CIDEGENT PlaGenT, Grant CIDEXG/2022/17, Project iTENTE, and the action CNS2023-144333, financed by MCIN/AEI/10.13039/501100011033 and the European Union “NextGenerationEU”/PRTR.} 

% \vspace{-3em}

}

\maketitle

%-- Abstract
\begin{abstract}\label{sec:Abstract}Energy consumption remains a dominant operational challenge for current and future cellular systems, especially in dense urban deployments.
This paper investigates a novel role for \ac{NTN} \ac{HAPS} as an enabler of energy-efficient operation rather than only coverage extension.
We define the \ac{HAPS}-Hypercell as a wide-area non-terrestrial layer that can assume the coverage role of multiple terrestrial macro-cells, enabling, for the first time, the shutdown of both capacity and coverage macro-cells. 
We develop a comprehensive \ac{3GPP}-compliant system model, along with two \ac{HAPS}-Hypercell pairing architectures that capture the interplay among multiple layers, realistic channel conditions, and distributed \ac{CS} mechanisms. 
Our results show that the \ac{HAPS}-Hypercell can effectively reduce overall network power consumption.
We then identify key limitations of a straightforward \ac{HAPS} integration, laying the groundwork for future optimization and providing key insights for next-generation \ac{CS} operations.
\end{abstract}
\glsresetall

%--- Introduction
\section{Introduction}\label{sec:Introduction}

The transition toward 5G-Advanced and emerging 6G systems is intensifying the energy challenge for mobile networks.
Driven by traffic growth, dense deployments, \ac{mMIMO} and advanced radio technologies,
\ac{RAN} energy consumption has become a major concern for operators and regulators.
Mobile networks consume 300–320\,TWh of electricity annually,
representing 1–1.3\% of global consumption,
while energy costs account for 20–40\% of operator operational expenditure~\cite{GSMA_climateAction24, IEA_2023}.
The environmental impact of the broader \ac{ICT} sector is also increasing,
with studies warning of a significant rise in greenhouse gas emissions if left unaddressed \cite{BELKHIR2018448}.
Improving network energy efficiency is therefore essential for the sustainable evolution of future wireless systems.

A key source of inefficiency lies in network design and operation. Operators distinguish between \emph{coverage cells}, typically high-power macro-cell that remain continuously active to ensure ubiquitous service, and \emph{capacity cells}, which are dynamically activated to accommodate traffic variations
and can leverage energy-saving mechanisms such as \ac{CS} during low-load conditions~\cite{10547043}.
However, this separation limits energy savings.
Coverage cells must remain active even at low traffic,
and their static power consumption could be high regardless of load.
As a result, energy optimization is largely restricted to the capacity layer,
leading to inefficient operation during off-peak periods.

In parallel, 5G and beyond architectures increasingly incorporate \acp{NTN},
including %\ac{LEO}
satellites and \acp{HAPS},
to extend network capabilities beyond terrestrial deployments \cite{9773096}.
Current efforts mainly focus on coverage enhancement,
targeting rural connectivity and service continuity. However, this perspective underutilizes the potential of \ac{NTN} for network optimization.

In this paper, we position \ac{HAPS} as an enabler of energy-efficient network operation.
Building on prior work on \ac{NTN} Hypercells~\cite{10571153},
we define a \ac{HAPS}-Hypercell as a wide-area layer capable of assuming the coverage role of multiple terrestrial macro-cells.
This enables the coordinated shutdown of both capacity and coverage layers,
reducing energy consumption while maintaining service availability.
Recent works have explored \ac{HAPS}-based traffic offloading and energy savings~\cite{10571153, 10299797, 10304250, 10578559, 11352980}.
However, they rely on simplified models and centralized optimization,
and do not capture multi-layer interactions, pairing mechanisms, and operational constraints.
They also focus primarily on partial infrastructure deactivation.

In contrast, this paper considers the coordinated shutdown of both capacity and coverage macro-cells enabled by \ac{HAPS}-Hypercells.
We propose a practical algorithm for 
\ac{CS} that builds on existing mechanisms currently implemented in operational networks, and extends them to support Hypercell-based coordination.
The proposed approach explicitly accounts for
multi-layer interactions, enabling realistic and deployable operation in real networks.
We further develop a detailed \ac{3GPP}-compliant system model, capturing channel conditions, traffic dynamics, and distributed \ac{CS} decision-making.
This enables a comprehensive system-level evaluation of the proposed approach.
The main contributions of this paper are summarized as follows: \emph{(i)} We formalize the \ac{HAPS}-Hypercell concept as an energy-efficient framework enabling coordinated terrestrial resource deactivation. \emph{(ii)} We introduce structured pairing mechanisms and practical \ac{CS} algorithms for \ac{HAPS}-aided networks under realistic operational constraints. \emph{(iii)} We perform detailed \ac{3GPP}-compliant system-level simulations to quantify energy savings and analyze performance trade-offs.

%--- Hypercell
\section{Hypercell Concept and Architecture}\label{sec:Hypercell}

The Hypercell concept is introduced to overcome the legacy structural limitation of terrestrial networks, enabling deactivation of both capacity and coverage layers.
The key design principle is to decouple the coverage function from the terrestrial infrastructure and delegate it to a wide-area layer with significantly lower energy cost per unit area.

\subsection{Hypercell Definition}

A Hypercell is a logical coverage entity supported by a wide-area \ac{NTN} platform 
which ensures \ac{QoS} and service continuity independently of the terrestrial network.
Formally, let $\mathcal{C}^{\rm tn}$ denote the set of terrestrial cells and $\mathcal{C}^{\rm tn}_{\rm on} \subseteq \mathcal{C}^{\rm tn}$ the subset of terrestrial active cells. 
Let $\mathcal{A}_c$ and $\mathcal{A}_H$ denote the coverage areas of terrestrial cell $c$ and the Hypercell, respectively. 
The coverage condition is given by
\begin{equation}
\mathcal{A}_{\rm tot}
\subseteq
\left( \bigcup_{c \in \mathcal{C}^{\rm tn}_{\rm on}} \mathcal{A}_c \right)
\cup
\mathcal{A}_H.
\end{equation}
Where $\mathcal{A}_{\rm tot}$ represents the total covered area.
This ensures that the target service area remains fully covered even when a large fraction of terrestrial cells is deactivated, enabling aggressive energy savings without compromising service availability.

\subsection{Architectural Extension of Carrier Shutdown}

The Hypercell extends existing \ac{CS} mechanisms by generalizing the role of network layers in the shutdown process.

In conventional operation, cells are statically classified into capacity and coverage layers, with only capacity cells eligible for shutdown.
With the introduction of the Hypercell, this classification becomes relative rather than fixed.
Specifically:
\begin{itemize}
\item Capacity cells follow conventional \ac{CS} operation and can be dynamically deactivated.
\item Coverage cells also become eligible for deactivation, as the Hypercell assumes their coverage role, effectively reducing them to capacity nodes.
\item In extended hierarchical configurations, coverage cells can act as intermediate nodes: they provide coverage to paired capacity cells while remaining eligible for shutdown thanks to the Hypercell layer.
\end{itemize}
This results in a multi-layer architecture in which cell roles are determined by their position in the hierarchy, rather than by predefined categories.
This expanded design space enables a broader set of feasible shutdown configurations under the same operational constraints.

\subsection{Hypercell Pairing Structures}

Efficient and realistic \ac{CS} relies on distributed decision-making, where each cell determines its state based on local and neighboring conditions.
Since a cell must ensure that its traffic can be absorbed when switching off, local information alone is insufficient. 
Pairing mechanisms address this by enabling structured information exchange: each capacity cell is paired with a predefined coverage cell with which it exchanges information and control signalling.
% This restricts the decision space and yields stable, locally implementable decisions.

Formally, let $\mathcal{C}^{\rm tn}_{\rm cov}$ and $\mathcal{C}^{\rm tn}_{\rm cap}$ denote the sets of terrestrial coverage and capacity cells, respectively, with $\mathcal{C}^{\rm tn}_{\rm cov} \cup \mathcal{C}^{\rm tn}_{\rm cap} = \mathcal{C}^{\rm tn}$. 
For each capacity cell $c \in \mathcal{C}^{\rm tn}_{\rm cap}$, let $\mathcal{P}_c$ denote its set of paired cells, and by $\mathcal{P}_c^{\rm cov} \subseteq \mathcal{C}^{\rm tn}_{\rm cov}$  the set of its paired coverage cells.
Here, we assume $|\mathcal{P}_c^{\rm cov}| = 1$, i.e., each capacity cell is paired with exactly one coverage cell, whereas a coverage cell may serve multiple capacity cells.

% \begin{figure}[!t]
%     \centering
%     \includegraphics[width=0.6\columnwidth]{Figures/PNG/hierarchical_architecture_both.png}
%     \vspace{-0.5em}
%     \caption{Illustration of the proposed \ac{HAPS}-Hypercell pairing architectures.}
%     \label{fig:haps_pairing_architectures}
%     \vspace{-1.5em}
% \end{figure}
\begin{figure}[!t]
    \centering
    \includegraphics[width=0.75\columnwidth]{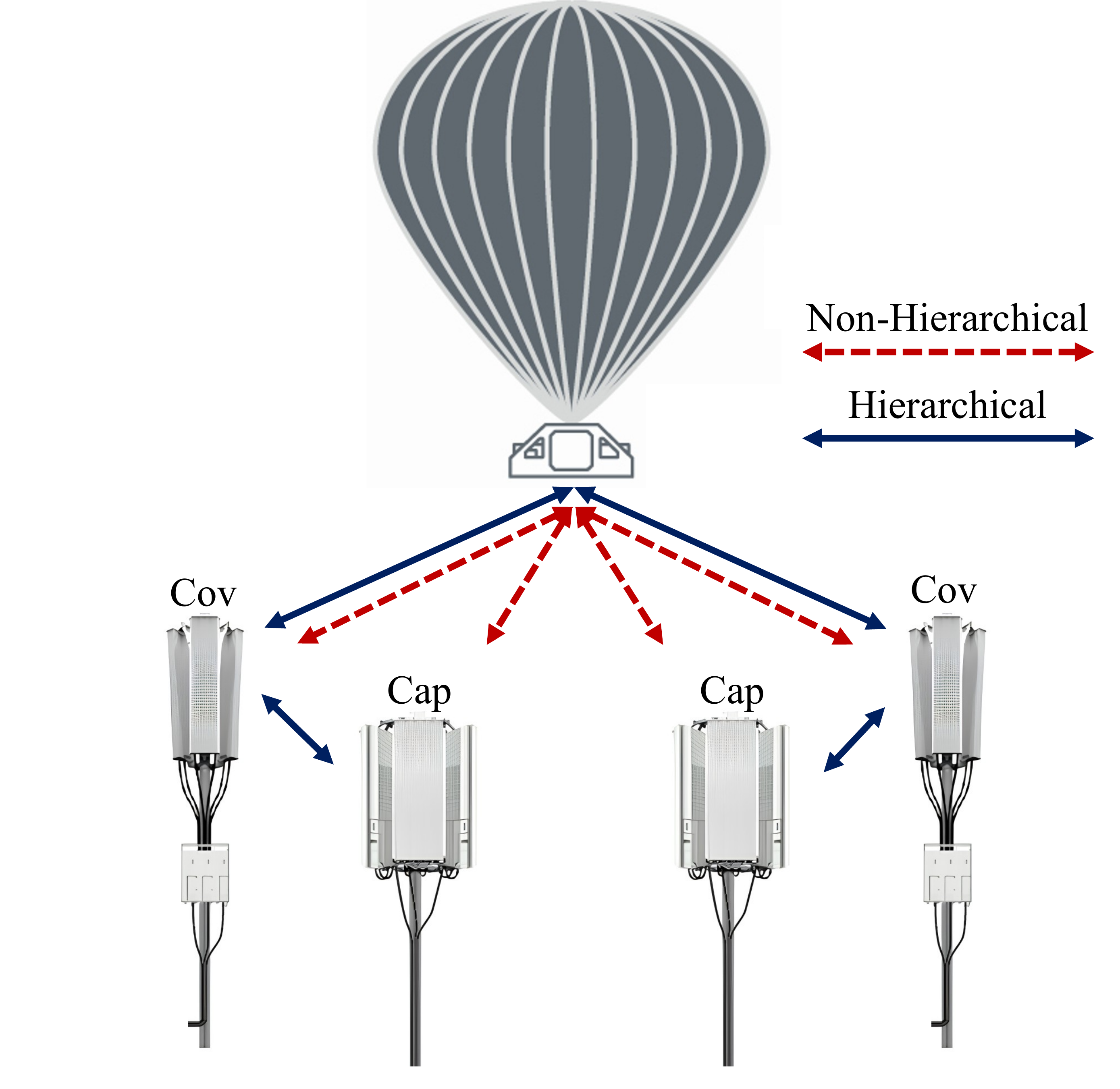}
    \vspace{-0.5em}
    \caption{Illustration of the proposed \ac{HAPS}-Hypercell pairing architectures.}
    \label{fig:haps_pairing_architectures}
    \vspace{-1.5em}
\end{figure}

In the Hypercell framework, legacy pairing is extended to include the \ac{NTN} layer.
Depending on the architecture, terrestrial cells may be paired either directly with the Hypercell or through intermediate layers, enabling multi-layer coordination. % while preserving the same distributed decision principles.
We consider two pairing architectures, illustrated in Figure~\ref{fig:haps_pairing_architectures}:
\subsubsection{Non-Hierarchical Architecture (HAPS-NH)}
    In the non-hierarchical configuration, the Hypercell acts as the sole coverage layer, and all terrestrial cells are treated as capacity nodes. Resulting in each cell paired directly with the Hypercell, 
    \begin{equation}
        \mathcal{C}^{\rm tn}_{\rm cap} = \mathcal{C}^{\rm tn}, \; \mathcal{C}^{\rm tn}_{\rm cov} = \emptyset \;\;\;
        \mathrm{and}\;\;\; \forall c \in \mathcal{C}^{\rm tn}_{\rm cap}, \;  \mathcal{P}_c^{\rm cov} = \{\mathcal{H}\} \;.
    \end{equation}
    This allows all terrestrial cells to directly exchange information and evaluate shutdown with respect to the Hypercell, maximizing flexibility and energy savings.

\subsubsection{Hierarchical Architecture (HAPS-H)}
In the hierarchical configuration, pairing relationships follow a multi-layer structure: terrestrial capacity cells $\mathcal{C}^{\rm tn}_{\rm cap}$ are paired with terrestrial coverage cells, while terrestrial coverage cells $\mathcal{C}^{\rm tn}_{\rm cov}$ are paired with both terrestrial capacity cells and the Hypercell $\{\mathcal{H}\}$.
This leads to the following pairing relations:
\begin{align}
    \forall c \in \mathcal{C}^{\rm tn}_{\rm cap},\, \mathcal{P}_c^{\rm cov} \subseteq \mathcal{C}^{\rm tn}_{\rm cov} \\ 
    \forall b \in \mathcal{C}^{\rm tn},\, \mathcal{P}_b^{\rm cov} \subseteq \mathcal{C}^{\rm tn}_{\rm cap} \cup \{\mathcal{H}\}
\end{align}
In the hierarchical architecture, the intermediate coverage cells $\mathcal{C}^{\rm tn}_{\rm cov}$ play a dual role: they are eligible for shutdown, thanks to Hypercell coverage and, when active, monitor their load and wake up deactivated paired capacity cells $\mathcal{C}^{\rm tn}_{\rm cap}$, if needed.
This pairing architecture preserves compatibility with legacy \ac{CS} mechanisms while extending them to multi-layer operation.

The two architectures exhibit trade-offs in flexibility and robustness.
The non-hierarchical configuration is fully flexible, as all cells interact directly with the Hypercell, but suffers from two limitations: the Hypercell becomes a global bottleneck under high load, and  reactivation is poorly targeted, e.g., the Hypercell may activate a cell that is not best positioned to absorb the traffic.
The hierarchical architecture introduces a coverage layer that enables localized control, allowing shutdown even when the Hypercell is highly loaded and improving reactivation accuracy.
However, it depends on intermediate decisions: poor load distribution at the coverage layer can create bottlenecks, e.g., a capacity cell may remain active despite available Hypercell capacity.

%--- Hypercell
\subsection{Hypercell-Based Carrier Shutdown Algorithm}\label{subsec:algorithm}

We consider a distributed \ac{CS} algorithm where each cell autonomously decides, based on local and paired-cell information, whether to deactivate itself or reactivate its paired cell.

Following practical implementations, the algorithm operates over discrete decision intervals, to iteratively reduce the number of active nodes and consolidate traffic into a smaller active set, thus switching the maximum number of cells to sleep mode while preserving \ac{QoS} and service continuity.

\subsubsection{Iterative Decision Process}
Each decision interval is divided into multiple steps:
\begin{itemize}
    \item \textit{Information Exchange}: each cell exchanges load information with its paired node.
    \item \textit{Shutdown Evaluation}: each shutdown eligible cell checks whether its shutdown condition is met and whether its load can be offloaded to the paired node.
    \item \textit{Wake-Up Evaluation}: each active coverage cell and the \ac{HAPS}-Hypercell monitor their respective loads and, if the wake-up conditions are met, send a wake-up trigger to a paired deactivated cell.    
    \item \textit{State Update}: each cell applies its deactivation or reactivation, transitioning to the new state.
\end{itemize}
To avoid oscillations, each coverage node can trigger at most one reactivation per decision step. 
In the analyzed scenarios, the system typically converges within tens of iterations.

\subsubsection{Shutdown and Wake-Up Operations}\label{subsec:shutdown_WakeUp_condition}
For each cell eligible for shutdown $c$, a shutdown decision is taken if its aggregate load with at least one of its paired coverage cells remains below a predefined threshold:
\begin{equation} \label{eq:shutdown_condition}
    \exists\, b \in \mathcal{P}_c^{\rm cov}
    \quad \text{s.t.} \quad
    \delta_c + \delta_b \leq \theta_{c,b}^{\rm shtdn},
\end{equation}
where $\delta_c$ and $\delta_b$ denote the \ac{PRB} loads of cells $c$ and $b$, respectively, and $\theta_{c,b}^{\rm shtdn} \in (0,1]$ is the shutdown threshold associated with the pair $(c,b)$. This condition provides a low-complexity approximation of traffic transfer feasibility while preserving distributed operation.
Conversely, a coverage cell $b$ triggers the reactivation of a paired deactivated node when its \ac{PRB} load exceeds a predefined threshold, i.e., $\delta_b > \theta_{b}^{\rm wkup}$.
When multiple paired cells are deactivated, the coverage cell prioritizes the one that served the largest number of \ac{UE} before shutdown.

In this work, $\theta^{\rm shtdn}$ and $\theta^{\rm wkup}$ are assumed identical across all capacity and coverage cells, respectively.

%--- System Model
\section{System Model}\label{sec:System_Model}

In this work, we consider a dense urban multi-layer network comprising co-located 4G, 5G cells and the \ac{HAPS}-Hypercell. For all layers, \acp{RAT} and \acp{RU}, we adopt models outlined by \ac{3GPP}~\cite{3GPP38901, 3GPP38811}.

\subsubsection{Network Layout}
The network consists of terrestrial and non-terrestrial layers.
The terrestrial segment includes co-located 4G \ac{LTE} and 5G \ac{NR} macro-cells operating at 2\,GHz and 3.5\,GHz, respectively. Both layers follow a \ac{3GPP} \ac{UMa} two-tier hexagonal grid deployment, with 19 tri-sectorized sites at a height of 25\,m and an \ac{ISD} of 500\,m, yielding 57 cells per layer and 114 cells in total. The two terrestrial layers are co-located, i.e., sites are geographically aligned.
The non-terrestrial segment comprises a single-sector \ac{HAPS} located at the network centre that covers the entire underlying area. The \ac{HAPS} is equipped with a 5G \ac{RU} operating at 1.8\,GHz.

In this work, 4G \acp{RU} transmit at 46\,dBm over a 20\,MHz bandwidth with 100 \acp{PRB} of 180\,kHz each, while 5G \acp{RU} transmit at 49\,dBm over a 100\,MHz bandwidth with 273 \acp{PRB} of 360\,kHz each, in accordance with \ac{3GPP} numerology.

The 5G layer acts as the capacity layer, the 4G layer as the coverage layer, and the \ac{HAPS} as the coverage Hypercell.

% \subsubsection{UE Distribution}
% \acp{UE} are evenly split between a uniform distribution over the network area and hotspots.
% Specifically, 15 circular hotspots with a radius of 40\,m are uniformly and randomly placed around the network center and the first-tier sites, with a minimum distance of 100\,m between their centers.
% %
% To evaluate performance under varying traffic densities, multiple hourly traffic conditions are considered. 
% Following the EARTH project~\cite{EARTH_D23_2010}, an average of 9 active \acp{UE} per cell is expected during peak hour, yielding up to 1026 active \acp{UE} across the 114-cell network. 
% Daily traffic variations are captured by applying the hourly proportions reported in~\cite{EARTH_D23_2010}; 
% \ac{UE} and hotspot positions are updated accordingly over time. 
% % Figure~\ref{fig:proportion_ue} shows the resulting traffic proportions and the corresponding number of active \acp{UE} throughout the day. 
% The set of all active \acp{UE} is denoted by $\mathcal{U}$.

\subsubsection{UE Distribution}
\acp{UE} are evenly split between a uniform distribution across the network, and 15 circular hotspots of radius 40\,m, randomly placed around the network center and first-tier sites with a minimum inter-hotspot distance of 100\,m.
To capture daily traffic variations, 
we follow the EARTH project~\cite{EARTH_D23_2010}, assuming an average of 9 active \acp{UE} per cell at peak hour, i.e., up to 1026 active \acp{UE} across the 114-cell network, and scale the number of active \acp{UE} according to the reported hourly traffic proportions.
\ac{UE} and hotspot positions are updated over time.
The set of active \acp{UE} is denoted by $\mathcal{U}$.

\subsubsection{Cell MIMO Antenna Array}
Each 4G cell employs a vertical \ac{ULA} with $N=8$ vertically polarized antenna elements connected to a single transceiver, while each 5G cell uses a \ac{UPA} with $M=32$ vertically polarized elements arranged in $M_v=4$ rows and $M_h=8$ columns, each connected to its own transceiver. 
The \ac{HAPS} is equipped with a downtilted \ac{UPA} analogous to 5G \acp{RU}, directed toward the ground. 
In all cases, the inter-element spacing is $\lambda_p/2$, with $\lambda_p$ equal to the carrier wavelength. 
Then, each \ac{UE} is equipped with a single antenna. 

Here, 4G cells operate with a fixed beam generated by the \ac{ULA}, whereas 5G cells and the \ac{HAPS} leverage digital precoding and \ac{mMIMO} capabilities through active antenna arrays. In this work, a single \ac{mMIMO}-layer per \ac{UE} is considered.

For each cell $c$, the matrix ${\bf V}_c = [{\bf v}_c^x, \, {\bf v}_c^y, \, {\bf v}_c^z]^T$ collects the Cartesian coordinates of its antenna elements w.r.t.  panel center; for the \ac{ULA} case, only ${\bf v}_c^z$ coordinates are non-zero.

\begin{figure*}[!tb]
\begin{equation}
\label{eq:SINR_comp}
% \scalebox{1.095}{% Ridimensiona l'equazione a 80% delle dimensioni originali
% $
\tag{10}
\gamma_{u,k} = 
    \frac{
    \beta_{u, \hat{c}_u} \left| {\bf h}_{u, \hat{c}_u, k}  {\bf w}_{u, \hat{c}_u, k} \right|^2 P_{u,\hat{c}_u, k}
    }
    {
    \beta_{u, \hat{c}_u}  \sum_{p \in \mathcal{U}_{\hat{c}_u} \setminus u}
     \left| {\bf h}_{u, \hat{c}_u, k}  {\bf w}_{p, \hat{c}_u, k} \right|^2   P_{p,\hat{c}_u, k} + 
    \sum_{c \in \mathcal{C} \setminus  \hat{c}_u} 
    \beta_{u, c}
    \sum_{i \in \mathcal{U}_c}
    \left| {\bf h}_{u, c, k}  {\bf w}_{i, c, k}\right|^2  P_{i,c,k} 
    +   
    \sigma_k^2
    }  
    % $
% }
\end{equation}
% \hrulefill
% \vspace*{4pt}
\vspace{-2.5em}
\end{figure*}

\subsubsection{Channel Model}
The channel between each \ac{UE} $u \in \mathcal{U}$ and each cell $c \in \mathcal{C}$ is modelled using the \ac{3GPP} \ac{UMa} statistical channel models from~\cite{3GPP38901} and~\cite{3GPP38811} for the terrestrial and \ac{NTN}, respectively.
The large-scale channel coefficient $\beta_{u,c}$ is computed as the product of the single-element antenna gain $g_{u,c}$, path loss gain $\rho_{u,c}$ and shadow fading gain $\tau_{u,c}$. Notably, $\tau_{u,c}$ is modelled as a spatially correlated random process, capturing realistic fading variations across \acp{UE}.

The small-scale fading, computed per antenna element, is modelled as a Rician channel~\cite{3GPP38901}. The resulting channel vector ${\bf h}_{u,c,k}$ for each \ac{PRB} $k$ is given by
\begin{equation}\label{eq:ComplexChannelRician}
    {\bf h}_{u,c,k} =
    \sqrt{\frac{K}{1+K}} \, {\bf h}^{\rm LoS}_{u,c,k}
    +
    \sqrt{\frac{1}{1+K}} \, {\bf h}^{\rm NLoS}_{u,c,k},
\end{equation}
with
\begin{equation}\label{eq:ComplexChannelLOS}
    {\bf h}^{\rm LoS}_{u,c,k} = e^{-j \frac{2\pi}{\lambda_c} d^{\rm 3D}_{u,c}}
    \,
    e^{j \frac{2\pi}{\lambda_c} \, {\bf k}_{u,c}^T(\phi_{u,c}, \theta_{u,c}) \, {\bf V}_c}, \,
% \end{equation}
% \begin{equation}\label{eq:ComplexChannelNLOS}
    {\bf h}^{\rm NLoS}_{u,c,k} \sim
    \mathcal{CN}({\bf 0}, {\bf I}),
\end{equation}
where $K$ is the Rician factor, $d^{\rm 3D}_{u,c}$ is the 3D distance between \ac{UE} $u$ and the panel center of cell $c$, and $\phi_{u,c}$ and $\theta_{u,c}$ are the corresponding \ac{UE} azimuth and zenith angles.
The wave vector ${\bf k}_{u,c}\left( \cdot, \cdot \right)$ represents the phase variation of a plane wave in 3D-orthogonal directions, as defined in 3GPP~\cite{3GPP38901}.

\subsubsection{Cell Association}\label{subsec:CellAssociation}

In the initial phase of cell discovery and association, each cell transmits \acp{RS} to cover its area. 
In 4G networks, cell access is based on always-on \acp{CRS}, transmitted through the \ac{ULA} panel using a fixed beam codeword ${\bf w}_{c}^{\rm crs}$ designed to produce a wide beam downtilted at 102$^\circ$. In contrast, 5G cells and the \ac{HAPS} rely on multiple beamformed \acp{SSB}, each generated through a \ac{2D-DFT} codeword ${\bf w}_{s,c}^{\rm ssb}$ applied to the associated \ac{UPA} panel, ensuring even spatial coverage. The \acp{SSB} are transmitted sequentially via a beam-sweeping procedure.
In compliance with the \ac{3GPP} \ac{FR1} specifications, each 5G cell transmits up to 8 \ac{SSB} beams.
Each \ac{UE} $u$ selects its serving cell $\hat{c}_u$ as the one providing the maximum \ac{RSRP}. 
\Ac{RSRP} for 4G is based on \acp{CRS} while for 5G and \ac{HAPS}, it is computed per \ac{SSB} as follows: 
\begin{equation}\label{eq:rsrp_4G}
    {\rm RSRP}^{\rm 4G}_{u,c} = 
    \mathbb{E}_k\left[
    \beta_{u,c} \left| {\bf h}_{u,c,k} {\bf w}_{c}^{\rm crs} \right|^2 P^{\rm Tx,\,crs}_{c}
     \right]
    \; ,
\end{equation}
\begin{equation}\label{eq:rsrp_5G}
    {\rm RSRP}^{\rm 5G,HAPS}_{u,s,c} = 
    \mathbb{E}_k\left[
    \beta_{u,c} \left| {\bf h}_{u,c,k} {\bf w}_{s,c}^{\rm ssb} \right|^2 P^{\rm Tx,\,ssb}_{s,c}
    \right] \;.
\end{equation}
The expectation is taken over multiple channel realizations across the resource elements assigned for reference signals.

\subsubsection{Data Transmission KPI}
Each cell $c$ transmits data to its associated \acp{UE} $\mathcal{U}_c$ by multiplexing across \acp{PRB} and beams.

For 5G cells and the \ac{HAPS}, to harness \ac{mMIMO} beamforming and multiplexing capabilities, we adopt a Type-I \ac{CSI-RS} Codebook scheme~\cite{DahlmanBook}. 
In compliance with the \ac{3GPP} \ac{FR1} specifications, each 5G cell transmits up to 32 \ac{CSI-RS} beams, each precoded by a codeword selected from a \ac{2D-DFT} codebook. Then, each \ac{UE} estimates the received power on the corresponding \ac{RS} and feeds back a set of indices characterizing the instantaneous channel conditions. 
Based on the reported \ac{PMI}, which identifies the beam yielding the largest received power, the 5G cell sets the precoding vector ${\bf w}_{u, \hat{c}_u, k}$ of \ac{UE} $u$ equal to the codeword associated with the indicated beam.
In contrast, owing to its limited \ac{mMIMO} capabilities, the 4G cell transmits a single wide beam, whose precoding vector ${\bf w}_{u, \hat{c}_u, k}$ is fixed, identical for all served \acp{UE}, and corresponds to the codeword previously used for \ac{CRS}.
Finally, in all cells, \acp{PRB} are fully reused across beams within the same cell. The number of \acp{PRB} assigned to each \ac{UE} is computed as outlined later in Section~\ref{subsubsec:RateDemand}.

We denote by ${\bf w}_{u,c,k}$ the precoding codeword assigned to \ac{UE} $u$ by cell $c$ on \ac{PRB} $k$ and by $P_{u,c,k}$ the corresponding transmit power. The data \ac{SINR} $\gamma_{u,k}$ is defined in eq.~\eqref{eq:SINR_comp}. 
Then, the achievable rate of \ac{UE} $u$ is computed as
\begin{equation}\label{eq:data_rate}
\setcounter{equation}{11}
    R_u = |{\mathcal K}_{u,\hat{c_u}}| \; B^{\rm prb}_{\hat{c_u}} \; \log_2 (1 + \tilde{\gamma}_{u}),
\end{equation}
where $|{\mathcal K}_{u,\hat{c_u}}|$ is the number of assigned \acp{PRB}, $B^{\rm prb}_{\hat{c_u}}$ is the \ac{PRB} bandwidth in cell $\hat{c_u}$, and $\tilde{\gamma}_{u}$ is the effective \ac{SINR}, derived from the per-\ac{PRB} \acp{SINR} via mutual information effective \ac{SINR} mapping.

\begin{figure*}
    % Legenda condivisa sopra
    \centering
    \ref{shared_legend_powerConsumption}\\\vspace{-0.9em} %[0.5em]
    % First subplot  $\mu_{lin}=1.0,\ \sigma_{lin}=3.00$}$
    \subfloat[\footnotesize{Low Traffic Demand}\label{subfig:PowerCons_LowTraffic}]{
        \setlength\fwidth{0.975\columnwidth}
        \setlength\fheight{0.6026\columnwidth}
        \begin{tikzpicture}
\definecolor{darkgray176}{RGB}{176,176,176}
\definecolor{green}{RGB}{0,128,0}
\definecolor{lightgray204}{RGB}{204,204,204}
\begin{axis}[
width=\fwidth,
height=\fheight,
legend cell align={left},
legend style={
  fill opacity=0.9,
  draw opacity=1,
  text opacity=1,
  draw=lightgray204,
  font=\footnotesize,
  legend columns=6
},
legend to name=shared_legend_powerConsumption,
tick pos=left,
x grid style={darkgray176, opacity=0.5, dashed},
xlabel={Hour of Day},
xlabel style={yshift=6pt, font=\small},
xmajorgrids,
xmin=-1.2, xmax=25.2,
xtick style={color=black},
xtick={0,2,4,6,8,10,12,14,16,18,20,22,24},
xticklabel style={rotate=45.0, font=\footnotesize},
xticklabels={
  0:00,
  2:00,
  4:00,
  6:00,
  8:00,
  10:00,
  12:00,
  14:00,
  16:00,
  18:00,
  20:00,
  22:00,
  24:00
},
y grid style={darkgray176, opacity=0.5, dashed},
ylabel={Avg. Network Power (kW)},
ylabel style={font=\small},
ymajorgrids,
ymin=54.25, ymax=65.75,
ytick={55, 57,...,65},
ytick style={color=black},
yticklabel style={font=\footnotesize}
]
\addplot [semithick, dashed, blue, mark=*, mark size=1.5, mark options={solid}]
table {%
0 60.1598640093584
1 59.0391834598143
2 57.8800570287823
3 57.110973019568
4 55.5990905752538
5 55.4281417534072
6 55.6731426152149
7 56.7282216512818
8 57.5427033653854
9 57.5505973705211
10 57.7182553120214
11 58.7272675808865
12 58.8325978929002
13 58.5974473629076
14 58.6475931952396
15 58.7931211134989
16 58.7914831352113
17 58.8235415582139
18 59.1339289839982
19 59.5290857255338
20 59.6575805500506
21 59.8634784869789
22 60.0835575188556
23 59.650301910587
};
\addlegendentry{HAPS-NH Cons.}
\addplot [semithick, dashed, green, mark=*, mark size=1.5, mark options={solid}]
table {%
0 57.6763268560786
1 58.9159031057913
2 58.3689822437246
3 57.1326830755769
4 55.6354685493607
5 56.1294589860359
6 55.3955659760911
7 55.9220985658922
8 57.0263270566939
9 56.8757620815514
10 56.8420461853503
11 57.8040675111372
12 58.2116742633778
13 57.5882414702334
14 58.0625109395185
15 58.2805312105972
16 58.247579772051
17 58.2011145616689
18 58.9043649853705
19 59.263190094741
20 59.1564282073337
21 59.1204404254514
22 59.4369133221505
23 58.3950973728258
};
\addlegendentry{HAPS-NH Bal.}
\addplot [semithick, dashed, red, mark=*, mark size=1.5, mark options={solid}]
table {%
0 57.0007975602348
1 57.3812953112601
2 57.5860249157507
3 58.3629930802285
4 55.7656580312946
5 56.1295635942815
6 55.4330027505714
7 55.4965461956857
8 56.3369537694453
9 56.8137265533446
10 57.1153602061191
11 57.2553669692674
12 57.5588333175976
13 57.4923988571921
14 57.5106081583419
15 57.5887894058941
16 57.6000252090174
17 57.4561645119983
18 58.3371919864931
19 58.7428156383394
20 58.4757601634262
21 59.0027442074218
22 58.7714692665337
23 57.4318964537778
};
\addlegendentry{HAPS-NH Aggr.}
\addplot [semithick, blue, mark=triangle*, mark size=1.5, mark options={solid}]
table {%
0 60.4228202596723
1 59.5287781479278
2 58.3661279688436
3 56.8994501755395
4 55.4888065617918
5 56.0651693647781
6 55.8122793287077
7 56.4581165615517
8 57.269116225612
9 57.4115985742091
10 57.9604858782211
11 58.9977623675226
12 59.131876036616
13 58.6807971419532
14 58.7261095554668
15 58.8965386746683
16 58.8897235291917
17 58.8804774256109
18 59.2516333379466
19 59.9581796498377
20 60.2767715280333
21 60.6583602634786
22 61.0836946699618
23 60.445695622293
};
\addlegendentry{HAPS-H Cons.}
\addplot [semithick, green, mark=triangle*, mark size=1.5, mark options={solid}]
table {%
0 58.3178394229073
1 57.7401168100117
2 57.1280772869546
3 56.757788346362
4 55.3719285408098
5 56.0064732558925
6 55.9822874949812
7 56.121462071959
8 56.8208861356138
9 57.043343797016
10 57.2532912588515
11 58.2514157313544
12 57.8340067764519
13 57.6769032635369
14 58.0727255727369
15 58.0382054540553
16 57.9360374166924
17 57.8234323978065
18 58.7803710473814
19 59.4559595089871
20 59.0240604919591
21 59.3132774178504
22 59.8111004728117
23 57.8436113484739
};
\addlegendentry{HAPS-H Bal.}
\addplot [semithick, red, mark=triangle*, mark size=1.5, mark options={solid}]
table {%
0 57.3060706400671
1 57.1068558094897
2 57.1492167046545
3 57.1941500036198
4 55.9227120771804
5 55.8850552756109
6 55.3622396628935
7 55.8443975334524
8 56.2129193508226
9 56.6925581650733
10 57.1222409074782
11 57.3832272311527
12 57.598301650373
13 57.6111586259999
14 57.7426837060291
15 57.8433760853766
16 57.9148673297563
17 57.7946851162909
18 58.5678688540695
19 58.9180210724511
20 58.6923984367925
21 58.8361814223367
22 58.7137679783224
23 57.5943360757985
};
\addlegendentry{HAPS-H Aggr.}
\addplot [semithick, black]
table {%
0 64.8563896623075
1 64.6864424173931
2 64.5101201009134
3 64.282839849621
4 64.1178276615341
5 64.1470649050335
6 64.1524908367832
7 64.1142196077803
8 64.2682033357342
9 64.3940171996891
10 64.5273884271344
11 64.6572318700016
12 64.6177873349428
13 64.5549236803452
14 64.8028255694588
15 64.7318715851406
16 64.8390154153446
17 64.8002572481732
18 64.9817800550838
19 65.0991223435541
20 65.025541662391
21 65.2057071469863
22 65.2543730481684
23 64.9517212732832
};
\addlegendentry{HAPS No CS}
\addplot [semithick, blue, dotted]
table {%
0 60.9680159115929
1 60.1249046238342
2 59.7922786848504
3 59.0371048087873
4 58.6632591293472
5 58.709313417065
6 58.7478831237871
7 58.6998322163342
8 58.9081359374442
9 59.1093904551425
10 59.2639102892318
11 59.4264516050377
12 59.5251376770495
13 59.5341987318395
14 59.7593199128467
15 59.7704175457754
16 59.7685658926763
17 59.8256058879731
18 60.0410973198372
19 60.3049271455445
20 60.4710745982606
21 60.6925641417303
22 60.9697299733995
23 60.5768247525174
};
\addlegendentry{TN Cons.}
\addplot [semithick, green, dotted]
table {%
0 59.9961422956207
1 59.5370278778591
2 59.2026153913019
3 58.8729418882885
4 58.5883762373486
5 58.5943698804814
6 58.6212140373387
7 58.6098016357063
8 58.7824353308438
9 59.0654532565433
10 59.2601966873763
11 59.4317078580775
12 59.4930201888957
13 59.4041119561035
14 59.635898991223
15 59.6436547866502
16 59.6360056897042
17 59.6391783743857
18 59.8640556049226
19 60.0241051553605
20 60.0466451022186
21 60.168397217508
22 60.353777838377
23 59.9329751709857
};
\addlegendentry{TN Bal.}
\addplot [semithick, red, dotted]
table {%
0 59.8220268678783
1 59.4227182798861
2 59.2123555625119
3 58.8627664203046
4 58.5931052675126
5 58.6001988106011
6 58.6236026812989
7 58.599692129123
8 58.7801673279006
9 59.0669733010608
10 59.2587402659017
11 59.4458747094232
12 59.4229051394659
13 59.3368399095971
14 59.570766029314
15 59.5670128402708
16 59.5905776115614
17 59.6046860199728
18 59.7296658545055
19 59.9412931993244
20 59.9577094527481
21 60.0588216779628
22 60.1509561339138
23 59.8279770574965
};
\addlegendentry{TN Aggr.}
\addplot [semithick, black, dotted]
table {%
0 64.259553898863
1 64.0660005136907
2 63.9579298724393
3 63.7578390624006
4 63.5710149691462
5 63.5854675223529
6 63.6038836930195
7 63.5960948361873
8 63.7186902338942
9 63.8513658156335
10 63.9390778720836
11 64.0544619634748
12 64.0610799459755
13 64.009246851488
14 64.2259537968795
15 64.2587211807191
16 64.2428711130778
17 64.2509038939476
18 64.3846916718483
19 64.5278836956104
20 64.4990202580531
21 64.6291837318381
22 64.6844125187715
23 64.4211434907019
};
\addlegendentry{TN No CS}
\end{axis}

\end{tikzpicture}
    }
    \hfill
    \subfloat[\footnotesize{High Traffic Demand}\label{subfig:PowerCons_HighTraffic}]{
        \setlength\fwidth{0.975\columnwidth}
        \setlength\fheight{0.6026\columnwidth}
        % This file was created with tikzplotlib v0.10.1.
\begin{tikzpicture}

\definecolor{darkgray176}{RGB}{176,176,176}
\definecolor{green}{RGB}{0,128,0}
\definecolor{lightgray204}{RGB}{204,204,204}

\begin{axis}[
width=\fwidth,
height=\fheight,
legend cell align={left},
legend style={draw=none, fill=none},
tick pos=left,
x grid style={darkgray176, opacity=0.5, dashed},
xlabel={Hour of Day},
xlabel style={yshift=6pt, font=\small},
xmajorgrids,
xmin=-1.2, xmax=25.2,
xtick style={color=black},
xtick={0,2,4,6,8,10,12,14,16,18,20,22,24},
xticklabel style={rotate=45.0, font=\footnotesize},
xticklabels={
  0:00,
  2:00,
  4:00,
  6:00,
  8:00,
  10:00,
  12:00,
  14:00,
  16:00,
  18:00,
  20:00,
  22:00,
  24:00
},
y grid style={darkgray176, opacity=0.5, dashed},
ylabel={Avg. Network Power (kW)},
ylabel style={font=\small},
ymajorgrids,
ymin=57.0806788795752, ymax=73.2186243016041,
ytick style={color=black},
yticklabel style={font=\footnotesize}
]
\addplot [semithick, dashed, blue, mark=*, mark size=1.5, mark options={solid}]
table {%
0 70.6936010308028
1 68.0899358919958
2 68.2407952593108
3 62.3040289107819
4 59.3573484996914
5 59.8358951107938
6 60.9609862370709
7 60.9392213247199
8 63.4359640704393
9 63.694527511835
10 65.0186867576996
11 68.800593231974
12 69.0822751686315
13 68.2400306038141
14 69.8776262265841
15 69.9331387967269
16 70.3378930861711
17 70.4244057885488
18 71.1151543981473
19 71.188016873916
20 71.3306605241378
21 71.912112903889
22 72.4156774357081
23 70.8555609319608
};
% \addlegendentry{HAPS - NO\_hierarchical - Conservative (10, 60)}
\addplot [semithick, dashed, green, mark=*, mark size=1.5, mark options={solid}]
table {%
0 70.6223636611144
1 65.4203563585142
2 65.9751045345147
3 60.590021316411
4 57.8142218533038
5 58.7017769586284
6 59.3274317320703
7 60.0523211795766
8 60.9901150611717
9 61.5958730146009
10 63.357161130313
11 65.3008199397483
12 64.9137459867476
13 63.3551966464518
14 64.6906640979368
15 65.9096679904221
16 68.0141327424367
17 69.2561525621573
18 70.7569879237652
19 71.2578676858703
20 71.1250880494396
21 71.9690092877984
22 72.4850813278755
23 70.2732508350055
};
% \addlegendentry{HAPS - NO\_hierarchical - Balanced (30, 70)}
\addplot [semithick, dashed, red, mark=*, mark size=1.5, mark options={solid}]
table {%
0 68.146051789093
1 63.8816791893084
2 64.3391942337671
3 60.3981167113104
4 58.979811113723
5 58.6926934645969
6 59.9848022687156
7 58.0291717106897
8 60.4278124340811
9 60.8083495028176
10 62.1420348992267
11 63.8822544421671
12 63.0390968211808
13 61.9366352867999
14 62.5873798771857
15 63.0213180491446
16 65.2090006911992
17 66.5440805722117
18 66.7802156180222
19 68.6376131848017
20 69.9616899555961
21 71.49499228218
22 72.4300065583666
23 68.5573191090028
};
% \addlegendentry{HAPS - NO\_hierarchical - Aggressive (50, 90)}
\addplot [semithick, blue, mark=triangle*, mark size=1.5, mark options={solid}]
table {%
0 70.4918583499511
1 68.5324555037459
2 68.3930230061014
3 63.139751589177
4 59.364887332793
5 59.7462323418735
6 60.6352331700483
7 61.0935091595668
8 64.4717748268882
9 63.6522579212149
10 66.3618149802406
11 69.5443948893169
12 68.355407784458
13 67.7215993283272
14 69.6326956629952
15 69.8456325652997
16 70.1758144920429
17 70.2425361011903
18 71.0058430017392
19 71.1773523589929
20 71.2662800462882
21 71.9045558830341
22 72.3317567078908
23 70.8271660513163
};
% \addlegendentry{HAPS - hierarchical - Conservative (10, 60)}
\addplot [semithick, green, mark=triangle*, mark size=1.5, mark options={solid}]
table {%
0 69.9681305775324
1 66.6958144118567
2 66.605644637382
3 60.9391400205889
4 59.1323100797493
5 59.6337035869518
6 59.6139906810402
7 59.8215765469073
8 61.2575641297618
9 62.8561066381692
10 64.7911509823004
11 68.4559522749145
12 66.1308752908388
13 66.4453728477954
14 67.8008077953497
15 68.6307404072801
16 69.2196786154508
17 69.1248268108805
18 70.184572892046
19 70.7515910583695
20 70.5609486566663
21 71.5775984850129
22 72.1875004060507
23 69.5790104593913
};
% \addlegendentry{HAPS - hierarchical - Balanced (30, 70)}
\addplot [semithick, red, mark=triangle*, mark size=1.5, mark options={solid}]
table {%
0 67.9791064292351
1 64.7974109501202
2 64.7640362117051
3 60.2809917757172
4 58.0829517056384
5 59.1593535661696
6 59.6763525543529
7 60.0510666552542
8 61.0365427430112
9 61.2631397609074
10 63.2498720097223
11 66.8076008426506
12 64.2785158432086
13 63.528973406736
14 65.6652166131337
15 66.4008311105409
16 67.3732538184562
17 67.4522218813995
18 68.3533185231884
19 68.8310367633501
20 69.2152062800924
21 70.3689456156572
22 71.36069044197
23 69.2020964357137
};
% \addlegendentry{HAPS - hierarchical - Aggressive (50, 90)}
\addplot [semithick, black]
table {%
0 70.6765625969887
1 69.7631970581194
2 68.799966817379
3 66.7317118004441
4 65.4021453941425
5 65.5331580568393
6 65.4561970248798
7 65.4377043694218
8 66.7982520504534
9 67.7738152072469
10 68.6694706341783
11 69.9251602253517
12 69.0253609867891
13 68.9764192704936
14 70.1875701067488
15 69.9035546034813
16 70.2749964045366
17 70.4323172600627
18 71.1024414609989
19 71.2694055449327
20 71.2579527249336
21 71.9099225070516
22 72.4443821600119
23 70.941799993825
};
% \addlegendentry{HAPS - No CS}
\addplot [semithick, blue, dotted]
table {%
0 69.319592931517
1 68.4757036715627
2 67.1194665618996
3 64.4104107141256
4 61.2568352123855
5 61.6030505763768
6 61.8019859340667
7 61.8342797926643
8 63.9531788440048
9 65.9780792700489
10 67.0269738051971
11 68.2376662035863
12 68.1328663109322
13 68.1867330552757
14 69.3233266923964
15 68.9280391860723
16 69.1449948503137
17 69.5060463580291
18 70.0202003222227
19 70.7613054031531
20 70.6226698383014
21 71.2240817672968
22 71.6250600985567
23 70.3204530415694
};
% \addlegendentry{TN - TN - Conservative (10, 60)}
\addplot [semithick, green, dotted]
table {%
0 69.042597633481
1 67.4677881391406
2 65.6914908017655
3 63.3701436872581
4 60.9750526855964
5 60.9979379192192
6 61.3006643688022
7 61.2777782628197
8 63.1365543658712
9 64.9386521281242
10 65.8017763628403
11 67.4362559855401
12 67.085134583199
13 67.0400798779765
14 68.4818476017197
15 68.1097898029963
16 68.4530667349656
17 68.873910766236
18 69.5096739522179
19 70.4186731059273
20 70.2710353259802
21 70.9908808726152
22 71.6558126762549
23 69.9738086190701
};
% \addlegendentry{TN - TN - Balanced (30, 70)}
\addplot [semithick, red, dotted]
table {%
0 68.3695532412131
1 66.1903053672988
2 64.7856701185384
3 62.4711633362054
4 60.5139229811826
5 60.6668695890664
6 60.8574006842572
7 60.8757471802114
8 62.3463653733173
9 63.841340150686
10 64.8295327799081
11 65.9095908671378
12 65.6414542845964
13 65.3184463264504
14 66.9551359041015
15 66.6264949581145
16 66.9373465365489
17 67.454349006772
18 68.1692081414302
19 69.0662347009659
20 68.9734603933533
21 69.8882886026223
22 70.5968512368282
23 68.7391259400129
};
% \addlegendentry{TN - TN - Aggressive (50, 90)}
\addplot [semithick, black, dotted]
table {%
0 69.5684519334813
1 68.7475143439412
2 67.8349123176933
3 66.3489540386637
4 64.7809526249846
5 64.8376633786361
6 64.9250240014195
7 64.8910681180775
8 66.0045783383488
9 67.2076129091601
10 67.8954722988765
11 68.7544708349546
12 68.5507289061725
13 68.4695050525526
14 69.426410525616
15 69.1720053405523
16 69.3669497157057
17 69.613322068131
18 70.156471371603
19 70.7804367192666
20 70.5707815743248
21 71.2481380504688
22 71.666834350586
23 70.3589153942347
};
% \addlegendentry{TN - No CS}
\end{axis}

\end{tikzpicture}
    }
    % \vspace{-0.5em}
    \caption{Average network power consumption over a 24-hour period under different \ac{CS} profiles and traffic demand.\label{fig:PowerCons}}
\vspace{-1.5em}
\end{figure*}
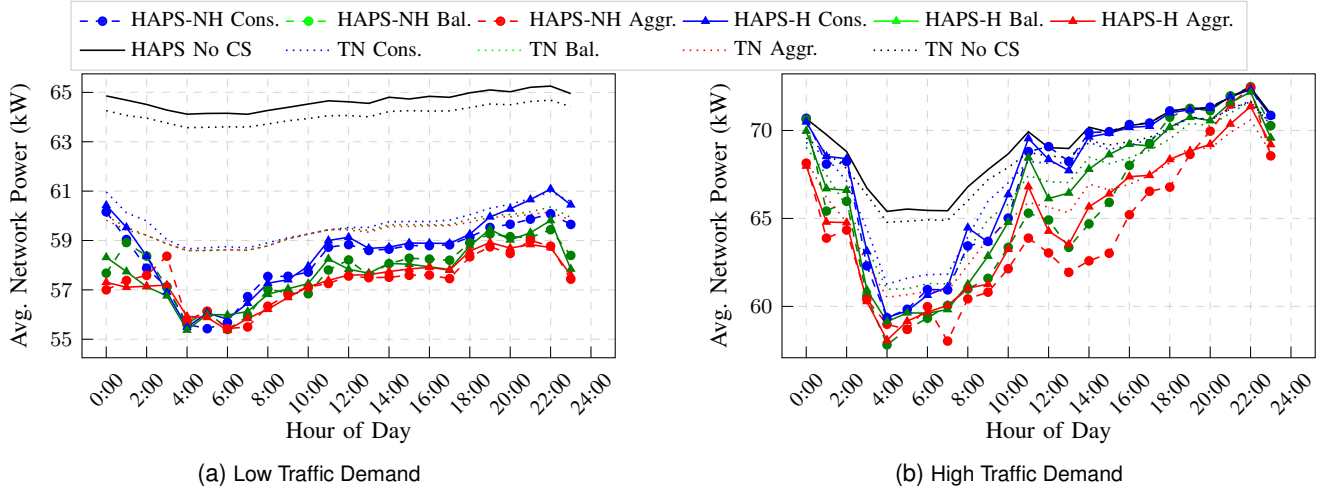

\subsubsection{Power Consumption}
Power consumption is modelled using the realistic single and multi carrier power model proposed in~\cite{9928089}:
\vspace{-1em}
\begin{equation}\label{eq:bs_power_model}
\resizebox{\columnwidth}{!}{$
P_{\rm BS} = P_{\rm BBU} + P_0 + P_{\rm BB}
+ \overbrace{M_{\rm TRX}^{\rm av} D_{\rm TRX}}^{P_{\rm TRX}}
+ \overbrace{M_{\rm PA}^{\rm ac} D_{\rm PA}}^{P_{\rm PA}}
+ \overbrace{\tfrac{1}{\eta} \sum_{c=1}^{C} P_{{\rm TX},c}}^{P_{\rm out}}
$}
\end{equation}
where $P_{\rm BBU}$ is the \ac{BBU} power consumption, $P_0$ is the power consumed by \ac{RU} interfaces and controllers, and $P_{\rm BB}$ accounts for baseband processing. $P_{\rm TRX} = M_{\rm TRX}^{\rm av} D_{\rm TRX}$ is the transceiver power, with $M_{\rm TRX}^{\rm av}$ active transceivers each consuming $D_{\rm TRX}$. Then, $P_{\rm PA} = M_{\rm PA}^{\rm ac} D_{\rm PA}$ is the static \ac{MCPA} power, with $M_{\rm PA}^{\rm ac}$ active \acp{MCPA} each consuming $D_{\rm PA}$. 

Each transceiver has a dedicated \ac{MCPA} that amplifies signals across all carriers and cells managed by the radio. $P_{\rm out}$ is the total radiated power, computed as the aggregate transmit power divided by the \ac{MCPA} and antenna efficiency $\eta$.

Numerical values of the consumption parameters in~\eqref{eq:bs_power_model} are based on commercial products and omitted for confidentiality.

\subsubsection{Rate Demands and Scheduling}\label{subsubsec:RateDemand}
Each \ac{UE} has a rate requirement $R^{\rm req}_u \sim \lognormal(\mu_{\rm req}, \sigma_{\rm req}^2)$, with parameters defined per traffic profile as given in Table~\ref{tab:traffic_profiles}.
The number of \acp{PRB} required to satisfy \ac{UE} $u$ requirement is computed as
\begin{equation}\label{eq:prb_demand}
    N^{\rm prb}_u  = \lceil {\frac{R^{\rm req}_{u}}{B^{\rm prb}_{\hat{c_u}}\; \log_2 (1 + \hat{\gamma}_{u})}} \rceil\;,
\end{equation}  
where $ \hat{\gamma}_{u}$ is the estimated \ac{SINR} of \ac{UE} $u$.

\subsubsection{Final Remarks}
At each \ac{CS} decision step, all stochastic components, channels and the network \acp{KPI} are recomputed, emulating the temporal evolution of a realistic, continuously varying environment.

All simulations are conducted using \texttt{Giulia}, a high-fidelity system-level simulator purpose-built for evaluating heterogeneous cellular networks through a combination of expert, \ac{3GPP}, and AI-based models. \texttt{Giulia} is publicly available at \url{https://github.com/giulia-open-lab/OpenGiuliaSLS.git}.

%--- Carrier Shutdown
% \input{Sections/05_Results_v1}
\section{Simulation Scenario and Results}\label{sec:SymSet_Results}

\begin{table}[t]
    \centering
    \caption{Traffic profile parameters for \ac{UE} rate demand.}
    \label{tab:traffic_profiles}
    \vspace{-0.75em}
    \begin{tabular}{lcccc}
        \hline
         & \textbf{Low} & \textbf{Medium} & \textbf{High} & \textbf{Very High} \\
        \hline
        $\boldsymbol{\mu_{\rm req}}$    & 0.0 & 1.60 & 2.70 & 3.0 \\
        $\boldsymbol{\sigma_{\rm req}}$ & 1.0 & 1.5  & 1.25 & 1.5 \\
        \hline
    \end{tabular}
    \vspace{-1.8em}
\end{table}
\begin{table}[t]
    \centering
    \caption{Threshold values for \ac{CS} profiles.}
    \label{tab:threshold_values}
    \vspace{-0.75em}
    \begin{tabular}{lccc}
        \hline
         & \textbf{Conservative} & \textbf{Balanced} & \textbf{Aggressive} \\
        \hline
        $\boldsymbol{\theta^{\rm shtdn}}$ & 0.10 & 0.30 & 0.50 \\
        $\boldsymbol{\theta^{\rm wkup}}$  & 0.60 & 0.70 & 0.90 \\
        \hline
    \end{tabular}
    \vspace{-2em}
\end{table}

This section presents the simulation scenario and discusses results in terms of power consumption and data rates, 
here representing network \ac{QoS}.
Here, we discuss the low and high-traffic scenarios, as they are the most representative of the conclusions drawn in this work.
We define three \ac{CS} policies parameterized by the thresholds $(\theta^{\mathrm{shtdn}}, \theta^{\mathrm{wkup}})$:
\begin{itemize}
\item \textit{Conservative}: low threshold values, 
resulting in limited shutdown and early reactivation,
\item \textit{Aggressive}: high threshold values, 
allowing increased shutdown and delayed reactivation,
\item \textit{Balanced}: intermediate threshold values,
providing a trade-off between energy savings and \ac{QoS}.
\end{itemize}
The parameter values are given in Table~\ref{tab:threshold_values}. For each setting, 30 iterations are considered to ensure \ac{CS} convergence.

\subsection{Power Consumption}
Figure~\ref{fig:PowerCons} shows the average network power consumption in kW for each hour and setting, averaged over all iterations and including both terrestrial cells and the \ac{HAPS}-Hypercell.

Enabling conventional \ac{CS} on \ac{TN} alone already yields savings of up to $7.7\%$--$7.8\%$ in the early-morning hours (4:00--7:00) under low traffic, 
depending on the profile,
confirming the effectiveness of legacy \ac{CS} methods.

Considering the \ac{HAPS}-Hypercell under low traffic with the conservative profile,
in the early-morning hours \textit{HAPS-NH} reduces consumption by $12.2\%$ relative to \textit{TN No CS} and by $4.8\%$ relative to \textit{TN Cons.}, 
while \textit{HAPS-H} achieves $12.0\%$ and $4.7\%$, respectively.
Even under conservative thresholds, 
these gains already match \ac{NGMN} reports and surpass those envisioned for future \ac{CS} solutions~\cite{ngmn2023nee}.
In terms of architectural pairing,
under these conditions, the gap narrows, since the \ac{HAPS} is far from saturation and both pairings can fully exploit the available flexibility 
(see Section~\ref{sec:Hypercell}). 
The slight \textit{HAPS-NH} advantage stems from pairing all terrestrial cells directly with the \ac{HAPS}, 
which prioritizes the direct reactivation of high-capacity 5G cells only,
whereas in \textit{HAPS-H},
the intermediate active cells may limit the achievable shutdown depth.
Switching to the aggressive profile, the gains extend across a broader portion of the day:
\textit{HAPS-NH} reaches $12.5\%$ in the early-morning hours and $10.5\%$ on a 24-hour average relative to \textit{TN No CS},
while \textit{HAPS-H} reaches $12.3\%$ and $10.5\%$, respectively. 

Under higher traffic (Figure~\ref{subfig:PowerCons_HighTraffic}), 
the same trends hold with reduced gains,
as more cells must remain active. 
The largest reduction relative to \textit{TN No CS} drops from $12.5\%$ to $9.2\%$ in the early-morning hours, 
both achieved by \textit{HAPS-NH Aggr.} 
Here, however, the architectural gap enlarges:
on a 24-hour basis, \textit{HAPS-NH Aggr.} saves $6.2\%$ versus $5.0\%$ for \textit{HAPS-H Aggr.}, outperforming it at every hour.
With all terrestrial cells paired directly to the \ac{HAPS} as capacity nodes,
the Hypercell reactivates 5G cells first, which, owing to their higher capacity, reduce the \ac{HAPS} load with fewer active cells.

\begin{figure*}[!t]
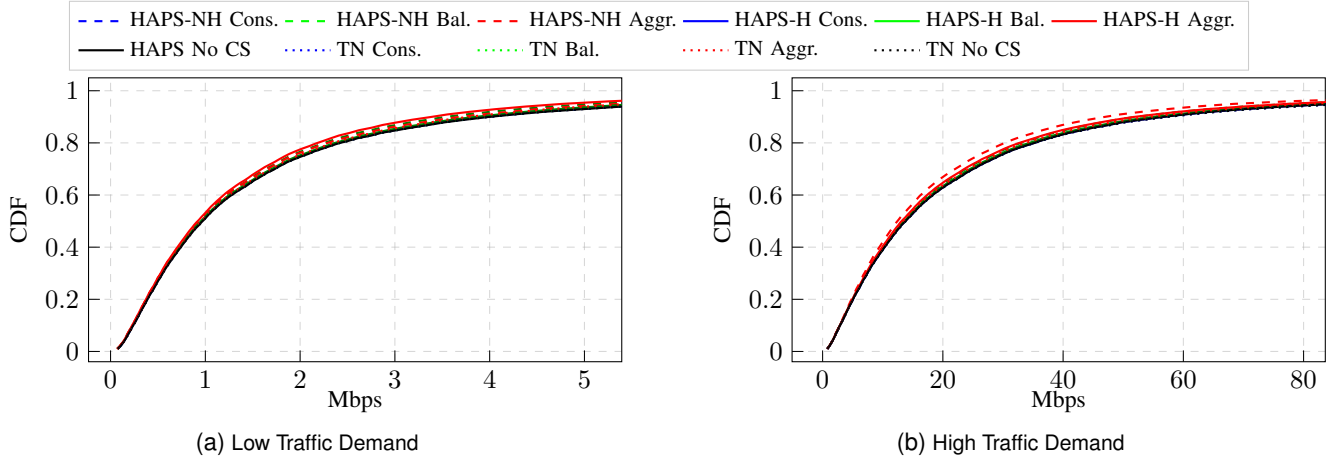

    % Legenda condivisa sopra
    \centering
    \ref{shared_legend_rate}\\\vspace{-0.9em} %[0.5em]
    % First subplot  $\mu_{lin}=1.0,\ \sigma_{lin}=3.00$}$
    \subfloat[\footnotesize{Low Traffic Demand}\label{subfig:RateDist_LowTraffic}]{
        \setlength\fwidth{0.975\columnwidth}
        \setlength\fheight{0.6026\columnwidth}
        \input{Figures/Tikz/Results/4G5G/Rate_dist/RateTraffic_mu_100_sigma_300}
    }
    \hfill
    \subfloat[\footnotesize{{High Traffic Demand}}\label{subfig:RateDist_HighTraffic}]{
        \setlength\fwidth{0.975\columnwidth}
        \setlength\fheight{0.6026\columnwidth}
        \input{Figures/Tikz/Results/4G5G/Rate_dist/RateTraffic_mu_1500_sigma_350}
    }
    % \vspace{-0.5em}
    \caption{Resulting \ac{UE} achievable rate distribution under different \ac{CS} profiles and traffic demand profiles.\label{fig:RateDist}}
\vspace{-1.5em}
\end{figure*}

\subsection{Rate Distribution}
We now analyze the rate distribution to assess \ac{QoS} impact. Figure~\ref{fig:RateDist} shows the rate \ac{CDF} for all configurations,
with the \textit{TN No CS} serving as reference. 

Under low traffic with the conservative profile,
both \textit{HAPS-NH} and \textit{HAPS-H} follow the no-\ac{CS} curve very closely, 
with mean rate reductions of only $1.3\%$ for both. 
The conservative thresholds restrict the number of deactivations and leave sufficient terrestrial capacity active, 
so the \ac{HAPS}-Hypercell does not noticeably affect \ac{QoS},
and the two architectures behave identically since the \ac{HAPS} is far from saturation.
As expected, the aggressive profile produces the largest impact, 
with mean rate reductions of $5.1\%$ for \textit{HAPS-NH} and $9.6\%$ for \textit{HAPS-H},
consistent with its prioritization of energy savings over \ac{QoS}. 
The balanced profile sits between the two ($7.0\%$ and $2.6\%$, respectively) and achieves the most favorable rate-energy trade-off, 
suggesting that intermediate threshold values are preferable in practice and that further fine-tuning could unlock additional gains.

Under higher traffic (Figure~\ref{subfig:RateDist_HighTraffic}), 
the rate curves deviate more noticeably from the no-\ac{CS} case,
especially for \textit{HAPS-NH} with the aggressive profile,
which suffers the largest mean rate reduction of $13.0\%$,
while \textit{HAPS-H Aggr.} is limited to $6.5\%$. 
This highlights a key trade-off between \ac{QoS}, power consumption, and pairing architecture.
\textit{HAPS-NH} enables the largest power savings due to its flexibility, but its control logic is concentrated at the \ac{HAPS} and relies solely on the local \ac{PRB} load: once the \ac{HAPS} becomes overloaded, it may reactivate suboptimal cells, leading to saturation and \ac{QoS} degradation.
\textit{HAPS-H} mitigates this issue: the \ac{HAPS} reactivates 4G cells, which in turn wake up paired 5G cells to further offload traffic.
This hierarchical control reduces saturation risk and improves \ac{QoS}, at the cost of higher power consumption. 
As traffic grows further, the overall benefits of the \ac{HAPS}-Hypercell solution diminish, as the network converges toward full cell activation.

%--- Conclusion
\section{Conclusion}\label{sec:Conclusion}

In this paper, we investigated the integration of a \ac{HAPS}-Hypercell into dense urban networks to reduce power consumption through \ac{CS}. 
For the first time, we showed that an \ac{NTN} \ac{HAPS} layer can assume the coverage role of multiple terrestrial macro-cells, enabling the joint shutdown of both capacity and coverage cells. 
Using a \ac{3GPP}-compliant multi-layer system model, we demonstrated power savings of up to 12.5\% during low-traffic hours. 
The results reveal that \ac{CS} decisions based solely on local \ac{PRB} load are insufficient to fully exploit the potential savings, and that the pairing architecture significantly impacts information sharing and achievable gains.

This work lays the foundation for \ac{HAPS}-enabled macro-cell shutdown, opening avenues for advanced algorithms, such as AI-based policies and reasoning models, to better capture network dynamics and maximize energy savings while guaranteeing \ac{QoS} toward future greener 6G networks.

%--- Bib
\bibliographystyle{IEEEtran}
\bibliography{journalAbbreviations, bibl}

%--- Author Bio
% \input{Utility/AuthorsBio}

\end{document}